\documentclass[sigconf]{acmart}

\renewcommand\footnotetextcopyrightpermission[1]{}
\acmConference[MM '23]{Proceedings of the 31st ACM International Conference on Multimedia}{October 29-November 3, 2023}{Ottawa, ON, Canada}
\acmBooktitle{Proceedings of the 31st ACM International Conference on Multimedia (MM '23), October 29-November 3, 2023, Ottawa, ON, Canada}
\acmDOI{10.1145/3581783.3613458}
\acmISBN{979-8-4007-0108-5/23/10}

\usepackage{hyperref}
\usepackage{multirow}
\usepackage{xspace}
\usepackage{soul}
\usepackage[ruled,vlined]{algorithm2e}
\SetAlFnt{\small}
\SetAlCapFnt{\small}
\SetAlCapNameFnt{\small}
\usepackage{cleveref}
\usepackage{pifont}
\newcommand{\cmark}{\ding{51}}
\usepackage{enumitem}
\usepackage{pgfplots}
\usepackage{makecell}
\usepackage{diagbox}
\usepgfplotslibrary{colorbrewer}
\usetikzlibrary{calc,shadings,patterns}

\pgfplotsset{compat=1.18}

\begin{document}

\title{\framework: A Unified Framework for the Extraction of Multimodal Features in Recommendation}

\author{Daniele Malitesta}
\authornote{Corresponding authors: Daniele Malitesta (\url{daniele.malitesta@poliba.it}) and Giuseppe Gassi (g.gassi@studenti.poliba.it).}
\affiliation{\institution{Politecnico di Bari, Italy}
  \city{}
  \country{}}
\email{daniele.malitesta@poliba.it}

\author{Giuseppe Gassi}
\authornotemark[1]
\affiliation{\institution{Politecnico di Bari, Italy}
  \city{}
  \country{}}
\email{g.gassi@studenti.poliba.it}

\author{Claudio Pomo}
\affiliation{\institution{Politecnico di Bari, Italy}
  \city{}
  \country{}}
\email{claudio.pomo@poliba.it}

\author{Tommaso {Di Noia}}
\affiliation{\institution{Politecnico di Bari, Italy}
  \city{}
  \country{}}
\email{tommaso.dinoia@poliba.it}

\renewcommand{\shortauthors}{Daniele Malitesta, Giuseppe Gassi, Claudio Pomo, \& Tommaso Di Noia}

\def\framework{\textsc{Ducho}\xspace}
\def\ducho{\textsc{Ducho}\xspace}


\keywords{Multimodal Recommendation, Deep Neural Networks}




\begin{abstract}
In multimodal-aware recommendation, the extraction of meaningful multimodal features is at the basis of high-quality recommendations. Generally, each recommendation framework implements its multimodal extraction procedures with specific strategies and tools. This is limiting for two reasons: (i) different extraction strategies do not ease the interdependence among multimodal recommendation frameworks; thus, they cannot be efficiently and fairly compared; (ii) given the large plethora of pre-trained deep learning models made available by different open source tools, model designers do not have access to shared interfaces to extract features. Motivated by the outlined aspects, we propose \framework, a unified framework for the extraction of multimodal features in recommendation. By integrating three widely-adopted deep learning libraries as backends, namely, TensorFlow, PyTorch, and Transformers, we provide a shared interface to extract and process features where each backend's specific methods are abstracted to the end user. Noteworthy, the extraction pipeline is easily configurable with a YAML-based file where the user can specify, for each modality, the list of models (and their specific backends/parameters) to perform the extraction. Finally, to make \framework accessible to the community, we build a public Docker image equipped with a ready-to-use CUDA environment and propose three demos to test its functionalities for different scenarios and tasks. The GitHub repository and the documentation are accessible at this link:~\url{https://github.com/sisinflab/Ducho}.
\end{abstract}

\maketitle

\section{Introduction and Motivation}

With the advent of the digital era and the Internet, numerous online services have emerged, including platforms for e-commerce, media streaming, and social networks. The vast majority of such websites rely on recommendation algorithms to provide users with a personalized surfing experience. In specific domains such as fashion~\cite{DBLP:conf/ecir/DeldjooNMM22}, music~\cite{DBLP:conf/recsys/OramasNSS17}, food~\cite{DBLP:journals/tmm/MinJJ20}, and micro-video~\cite{DBLP:conf/sigir/Yi0OM22} recommendation, recommender systems have demonstrated to be effectively supported in their decision-making process by all types of multimodal data sources the users usually interact with (e.g., product images and descriptions, users' reviews, audio tracks). 

The literature refers to multimodal-aware recommender systems (MRSs) as the family of recommendation algorithms leveraging multimodal (i.e., audio, visual, textual) content data to augment the representation of items, thus tackling issues in the field such as the sparsity of the user-item matrix and the inexplicable nature of users' actions (e.g., clicks, views) on online platforms which may not always be easy to profile for the recommendation algorithms.

Despite being the initial stage of any multimodal recommendation pipeline, the extraction of meaningful multimodal features is paramount in delivering high-quality recommendations~\cite{DBLP:conf/cvpr/DeldjooNMM21}. However, the current practice of employing diverse multimodal extraction procedures in each recommendation framework poses limitations. Firstly, these diverse implementations hinder the interdependence across various multimodal recommendation frameworks, making their fair comparison difficult~\cite{DBLP:conf/kdd/MalitestaCPDN23}. Secondly, despite the availability of numerous pre-trained deep learning models in popular open source libraries, the lack of shared interfaces for feature extraction across them represents a challenge for model designers.

To address these shortcomings, we propose \textbf{\framework}, a unified framework designed to streamline the extraction of multimodal features for recommendation systems. By integrating widely-adopted deep learning libraries as backends such as TensorFlow, PyTorch, and Transformers, we establish a shared interface that empowers users to extract and process audio, visual, and textual features from both items and user-item interactions (see \Cref{tab:overview}). This abstraction allows to leverage methods from each backend without being encumbered by the specific implementation that backend poses. A notable feature of our framework lays in its easily configurable extraction pipeline, which can be personalized using a YAML-based file. Users can specify the desired models, their respective backends, and models' parameters (such as the extraction layer). 

By looking at the related literature, the most similar application to \framework is Cornac~\cite{DBLP:journals/jmlr/SalahTL20}, a framework for multimodal-aware recommendation. For the sake of completeness, we report their main differences. Differently from Cornac, \framework: (i) is specifically aimed to provide customizable multimodal feature extractions, being completely agnostic to the downstream recommender system that might exploit the extracted features, thus being easily applicable to any model; (ii) provides the user with the possibility to select the deep learning extraction model, its backend, and its output layer; (iii) introduces the audio modality to the modalities set. 

To foster the adoption of \framework, we also develop a public Docker image pre-equipped with a ready-to-use CUDA environment\footnote{\url{https://hub.docker.com/r/sisinflabpoliba/ducho}.}, and propose three demos to show \framework's functionalities. The GitHub repository, which comes with all needed resources is available at:~\url{https://github.com/sisinflab/Ducho}.

\begin{table}[!t]
\caption{An overview of all modalities, sources, and backends combinations available in \framework.}
\label{tab:overview}
\centering
\footnotesize
\begin{tabular}{cccccc}
\toprule
\multirow{3}{*}{\textbf{Modalities}} & \multicolumn{2}{c}{\textbf{Sources}} & \multicolumn{3}{c}{\textbf{Backends}} \\ \cmidrule(lr){2-3} \cmidrule(lr){4-6}
& \multicolumn{1}{c}{Items} & Interactions & \multicolumn{1}{c}{TensorFlow} & \multicolumn{1}{c}{PyTorch} & Transformers \\ \cmidrule{1-6}
\textbf{Audio} & \multicolumn{1}{|c}{\cmark} & \multicolumn{1}{c|}{\cmark} &  & \multicolumn{1}{c}{\cmark} & \cmark \\
\textbf{Visual} & \multicolumn{1}{|c}{\cmark} &  \multicolumn{1}{c|}{\cmark} & \multicolumn{1}{c}{\cmark} & \multicolumn{1}{c}{\cmark} &  \\
\textbf{Textual} & \multicolumn{1}{|c}{\cmark} & \multicolumn{1}{c|}{\cmark} & \multicolumn{1}{c}{} & \multicolumn{1}{c}{} & \cmark \\ \bottomrule
\end{tabular}
\end{table}
\section{Architecture}

\framework's architecture is built upon three main modules, namely, \textbf{Dataset}, \textbf{Extractor}, and \textbf{Runner}, where the first two modules provide different implementations depending on the specific modality (i.e., audio, visual, textual) taken into account. We also remind the \textbf{Configuration} one among the other auxiliary components. The architecture is designed to be highly modular, possibly integrating new modules or customizing the existing ones. In the following, we dive deep into each outlined module/component.

\subsection{Dataset}
The \textbf{Dataset} module manages the loading and processing of the input data provided by the user. Starting from a general shared schema for all available modalities, this module provides three separate implementations: \textbf{Audio}, \textbf{Visual}, and \textbf{Textual} Datasets. As a common approach in the literature, the Audio and Visual Datasets require the path to the folder from which image/audio files are loaded, while the Textual Dataset works through a tsv file mapping all the textual characteristics to the inputs. 

Noteworthy, and differently from other existing solutions, \framework may handle each modality in two fashions, depending on whether the specific modality is describing either the \textbf{items} (e.g., product descriptions) or the \textbf{interactions} among users and items (e.g., reviews~\cite{DBLP:conf/cikm/AnelliDNSFMP22}). Concretely, while items are mapped to their unique ids (extracted from the filename or the tsv file), interactions are mapped to the user-item pair (extracted from the tsv file) they refer to. Although the pre-processing and extraction phases do not change at items- and interactions-level (see later), we believe this schema may perfectly suit novel multimodal-aware recommender systems with modalities describing every type of input source (even \textbf{users}).  

Another important task for the Dataset module is to handle the pre-processing stage of data input. Depending on the specific modality involved, \framework offers the possibility to:
\begin{itemize}[leftmargin=*]
    \item \textbf{audio:} load the input audio by extracting the waveform and sample rate, and re-sample it according to the sample rate the pre-trained model was trained on;
    \item \textbf{visual:} convert input images into RGB and resize/normalize them to align with the pre-trained extraction model;
    \item \textbf{textual:} (optionally) clean the input texts to remove or modify noisy textual patterns such as punctuation and digits.
\end{itemize}

After the extraction phase (see later), the Dataset module is finally in charge of saving the generated multimodal features into \textbf{numpy} array format following the file naming scheme from the previous mapping.
 
\subsection{Extractor}
The \textbf{Extractor} module builds an extraction model from a pre-trained network and works on each loaded/pre-processed input sample to extract its multimodal features. In a similar manner to the Dataset module, the Extractor provides three different implementations for each modality, namely, the \textbf{Audio}, \textbf{Visual}, and \textbf{Textual} Extractors. \framework exposes a wide range of pre-trained models from three main backends: TensorFlow, PyTorch, and Transformers. The following modality/backend combinations are currently available: 
\begin{itemize}[leftmargin=*]
    \item \textbf{audio:} PyTorch (Torchaudio) and Transformers;
    \item \textbf{visual:} Tensorflow and PyTorch (Torchvision);
    \item \textbf{textual:} Transformers (and SentenceTransformers).
\end{itemize}

To perform the feature extraction, \framework takes as input the (list of) extraction layers for any pre-trained model. Since each backend handles the extraction of hidden layers within a network differently, we follow the guidelines provided in the official documentations, assuming that the user will follow the same naming/indexing scheme of the layers and know the structure of the selected pre-trained model in advance. The interested reader may refer to the README\footnote{\label{readme}\url{https://github.com/sisinflab/Ducho/blob/main/config/README.md}.} under the \texttt{config/} folder on GitHub for an exhaustive explanation on how to set the extraction layer in each modality/backend setting.

Finally, for the textual case, the user can also specify the specific task the pre-trained model should be trained on (e.g., sentiment analysis), as each pre-trained network may come with different versions depending on the training strategy.

\subsection{Runner}
The \textbf{Runner} module is the orchestrator of \framework, whose purpose is to instantiate, call, and manage all the described modules. With its API methods, this module can trigger the complete extraction pipeline (see later) of one single modality or all the modalities involved simultaneously. 

The Runner module is conveniently customized through an auxiliary \textbf{Configuration} component which stores and exposes all parameters to configure the extraction pipeline. Even if a default configuration is already made available for the user's sake, \framework allows to override some (or all) its parameters through an external configuration file (in YAML format) and/or key-value pairs as input arguments if running the scripts from the command line. Once again, we suggest the readers refer to the README under the \texttt{config/} folder on GitHub to understand the general schema of the YAML configuration file. 

\begin{figure*}[!t]
\centering
    \includegraphics[width=0.6\textwidth]{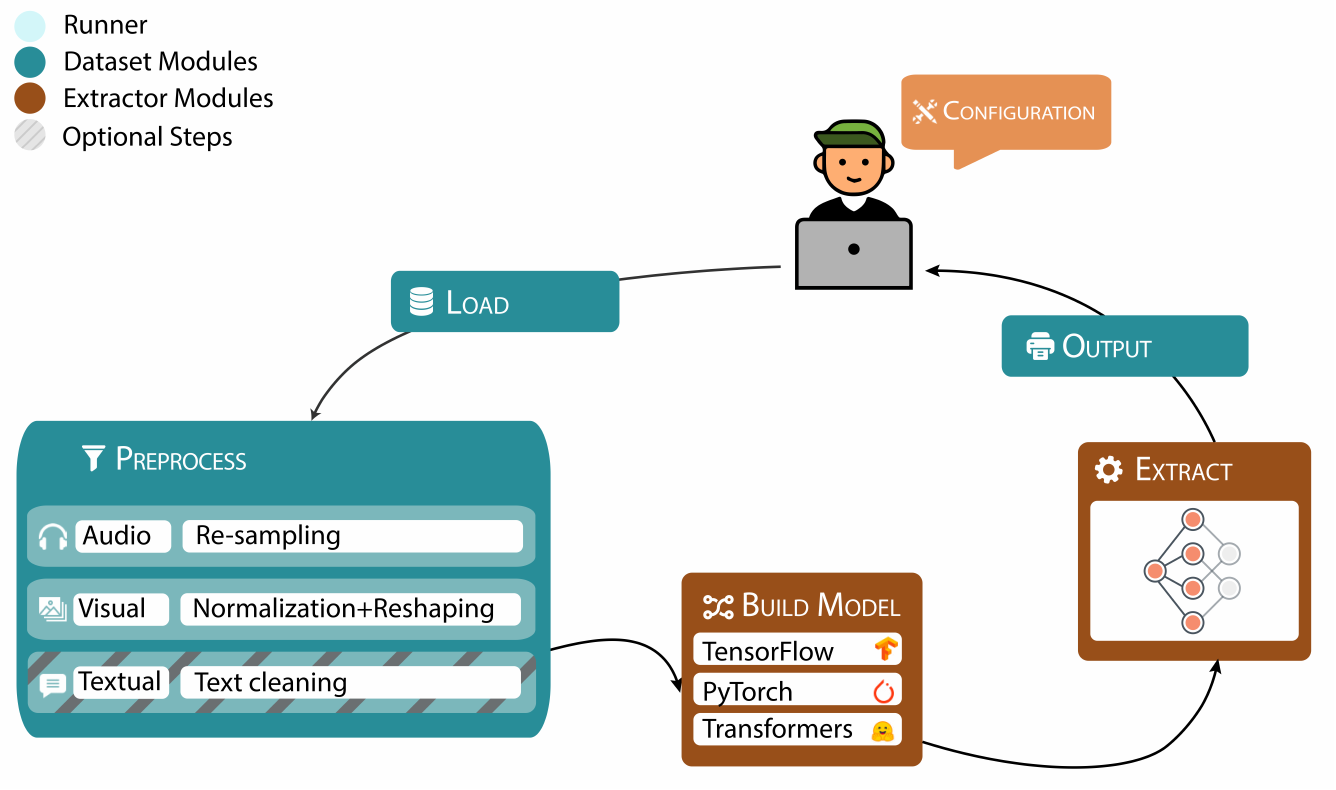}
    \caption{\framework's pipeline for multimodal feature extraction, managed by the Dataset, Extractor, and Runner modules.}
    \label{fig:framework}
\end{figure*}

\section{Extraction pipeline}

The overall multimodal extraction pipeline is represented in~\Cref{fig:framework}. Through the Dataset module, the \textbf{load} and \textbf{preprocess} steps take place, assuming that the user is providing the input data and the YAML configuration file (overridable from command line) to customize the extraction. Then, the Extractor module is in charge of \textbf{building} the extraction model(s) by setting the backends and output layer(s). Finally, after the multimodal feature \textbf{extraction}, features are saved to the \textbf{output} path in numpy format (the Dataset module again controls this latter phase). As previously described, the whole process is orchestrated by the Runner module.
\section{\framework as Docker application}

To fully exploit the GPU-speedup implemented in all backends we use for the multimodal feature extraction, one of the basic requirements is to setup a suitable development environment where the backends' versions are compatible with CUDA and, optionally, cuDNN. Generally, setting a workstation where all such libraries/tools are correctly aligned is challenging. To this end, we decide to dockerize \framework by making it into a Docker image (available on Docker Hub\footnote{\url{https://hub.docker.com/r/sisinflabpoliba/ducho}.}) with all packages already installed in a tested and safe virtualization environment on your physical machine. 

Our Docker image is built from an NVIDIA-based image which comes with CUDA 11.8 and cuDNN 8 on Ubuntu 22.04, Python 3.8 and Pip, and our cloned repository having all Python packages already installed and ready to be used. A possible container instantiated from the image should specify the gpus to use from the host machine (this feature is currently available on Docker but it depends on the version of CUDA to be installed), and the volume you may want to use to save the framework's outputs permanently. 

Note that a generic container instantiated from our image would prompt the user to a shell environment where one could run custom multimodal feature extractions via the command line, and also create custom configuration files for the same purpose.
\section{Demonstrations}
This section proposes three use cases (i.e., demos) which show some of the main functionalities in \framework and how to exploit them within a complete multimodal extraction pipeline. The guidelines and codes are accessible at this link\footnote{\url{https://github.com/sisinflab/Ducho/tree/main/demos}.} to run the demos (i) on your local machine, (ii) in a Docker container, and (iii) on Google Colab. Note that we specifically selected these demos as to replicate some real recommendation tasks involving multimodal features.

\subsection{Demo 1: visual + textual items features}
Fashion recommendation is probably one of the most popular task involving multimodal features to describe items. Generally, fashion products come with images (i.e., visual) and descriptions (i.e., textual) which may captivate the attention of the customer. 

\noindent\textbf{Input data.}
We use a small fashion dataset where each item has its own image and other metadata such as gender, category, colour, season, and product title. As for the visual modality, we save a subsample of 100 random images from the dataset in jpeg format; as for the textual modality, we produce for each of these items a description obtained as the combination of all the metadata fields from above, and store it into a tsv file where the first and second columns map item ids and descriptions, respectively. Note that, if no item column name is provided, \framework selects, by default, the last column as the one holding the items' descriptions. 

\noindent\textbf{Extraction.} In terms of extraction models, we adopt VGG19 and Xception for the product images, and Sentence-BERT pre-trained for semantic textual similarity for the descriptions. For each extraction model, we select the extraction layer, the pre-processing procedures, and the library where the deep network should be retrieved from.

\noindent\textbf{Output.} Through the configuration file, we set \framework to save the visual and textual embeddings to custom folders, where each embedding is a numpy array whose filename corresponds to the item name from the original input data. Additionally, \framework keeps track of the log file in a dedicated folder within the project.

\subsection{Demo 2: audio + textual items features}

When it comes to recommending songs to users, audio and textual features may enhance the representation of each song, where the former are structured as a waveform, the latter as sentences referring, for instance, to the music genre of the song. 

\noindent\textbf{Input data.} We use a small music genres dataset where each item comes with the binary representation of its waveform (we save it as wav audio track) and its music genre (we interpret it as textual song description and save it into a tsv file similarly to the previous demo). Given the heavy computational costs deep learning-based audio extractors require, we decide to select a small subset of the input songs (i.e., 10) just for the purpose of this demo.

\noindent\textbf{Extraction.} For the extraction of audio features we exploit Hybrid Demucs pre-trained for the task of music source separation. As for the textual extraction, we re-use the same deep neural model from the previous demo, since we are not interested in extracting other specific high-level features from music genres. 

\noindent\textbf{Output.} Once again, we use the configuration file to specify the output folders for both the audio and textual embeddings. Please note that the extraction of audio features might take some time depending on the machine you are running \framework on, as the deep audio extractor might require high computational resources to run. 

\subsection{Demo 3: textual items/interactions features}


Online platforms usually allow customers to express reviews and comments about the products they have enjoyed to share their experience with other potentially-interested customers. In an e-commerce scenario, items may come with textual descriptions of the product characteristics (as seen in Demo 1). However, textual reviews of users commenting on those items may also be involved. Unlike most existing literature works, which usually refer to both sources of information as items' representations, we decide to conceptually distinguish between items- and interactions (i.e., user-item)-side representations for the former and the latter, respectively.

\noindent\textbf{Input data.} We adopt the widely-popular Amazon recommendation dataset where each user's purchase keeps track of metadata such as customer/product ids, the review text, the rating, and the purchase date. In a similar manner to the other demos, we retain only a small subset of the original dataset including 100 reviews and the corresponding product descriptions (obtained as the concatenation of their product title and category). Specifically, we save descriptions and reviews into separate tsv files where the former follow the same format as Demo 1 and Demo 2, while the latter maps user/item ids to review texts. Note that the number of products does not correspond to the number of user-item interactions as we only consider the set of unique interacted items. While \framework extracts, by default, description/interaction texts from the last column of the tsv file, here we provide explicit column names to tell \framework where to retrieve product descriptions and user reviews from the respective tsv files.

\noindent\textbf{Extraction.} While for the items' descriptions we use again the same sentences encoder as in Demo 1 and 2, we decide to extract textual features from users' reviews through a multilingual BERT-based model pre-trained on customers' reviews and specify the task of sentiment analysis for this model.

\noindent\textbf{Output.} Textual item features are saved to numpy arrays whose filenames are the item ids. Conversely, the textual interaction features are saved under the filename obtained from user and item ids to provide a unique pointer to each review.
\section{Conclusion and Future Work}
In this paper we propose \textbf{\framework}, a framework for extracting high-level features for multimodal-aware recommendation. Our main purpose is to provide a unified and shared tool to support practitioners and researchers in processing and extracting multimodal features used as side information in recommender systems. Concretely, \framework involves three main modules: Dataset, Extractor, and Runner. The multimodal extraction pipeline can be highly customized through a Configuration component that allows the setup of the modalities involved (i.e., audio, visual, textual), the sources of multimodal information (i.e., items and/or user-item interactions), and the pre-trained models along with their main extraction parameters. To show how \framework works in different scenarios and settings, we propose three demos accounting for the extraction of (i) visual/textual items features, (ii) audio/textual items features, and (iii) textual items/interactions features. They can be run locally, on Docker (as we also dockerize \framework), and on Google Colab. As future directions, we plan to: (i) adopt all available backends (i.e., TensorFlow, PyTorch, and Transformers) to extract features for all modalities; (ii) implement a general extraction model interface allowing the users to follow the same naming/indexing scheme for all pre-trained models and their extraction layers; (iii) integrate the extraction of low-level multimodal features. 

\begin{acks}
This work was partially supported by the following projects: Secure Safe Apulia,
MISE CUP: I14E20000020001 CTEMT - Casa delle Tecnologie Emergenti Comune di Matera, CT\_FINCONS\_III, OVS Fashion Retail Reloaded, LUTECH DIGITALE 4.0, KOINÈ.
\end{acks}

\bibliographystyle{ACM-Reference-Format}
\bibliography{bibliography}

\end{document}